\documentclass[aps,pra,superscriptaddress,amsmath,amssymb,preprintnumbers,twocolumn,floats,showpacs]{revtex4}
\usepackage{amssymb}
\usepackage{graphicx}

\begin{document}

\title{Population trapping due to cavity losses}

\author{M. Scala}
\email{matteo.scala@fisica.unipa.it}

\affiliation{MIUR and Dipartimento di Scienze Fisiche ed
Astronomiche dell'Universit\`{a} di Palermo, via Archirafi 36,
I-90123 Palermo, Italy}

\author{B. Militello}

\affiliation{MIUR and Dipartimento di Scienze Fisiche ed
Astronomiche dell'Universit\`{a} di Palermo, via Archirafi 36,
I-90123 Palermo, Italy}

\author{A. Messina}

\affiliation{MIUR and Dipartimento di Scienze Fisiche ed
Astronomiche dell'Universit\`{a} di Palermo, via Archirafi 36,
I-90123 Palermo, Italy}

\author{S. Maniscalco}

\affiliation{Department of Physics, University of Turku, FI-20014
Turun yliopisto, Finland}

\author{J. Piilo}

\affiliation{Department of Physics, University of Turku, FI-20014
Turun yliopisto, Finland}

\author{K.-A. Suominen}

\affiliation{Department of Physics, University of Turku, FI-20014
Turun yliopisto, Finland}

\begin{abstract}
In population trapping the occupation of a decaying quantum level
keeps a constant non-zero value. We show that an atom-cavity
system interacting with an environment characterized by a non-flat
spectrum, in the non-Markovian limit, exhibits such a behavior,
effectively realizing the preservation of nonclassical states
against dissipation. Our results allow to understand the role of
cavity losses in hybrid solid state systems and pave the way to
the proper description of leakage in the recently developed cavity
quantum electrodynamic systems.
\end{abstract}


\maketitle

Population trapping arises when a multilevel system interacting
with external driving fields  is frozen in a given state for very
long times~\cite{YooandRadmorebrief,Deng,Sanchez1}. The phenomenon
was originally discovered in the dynamics of systems with at least
three discrete levels~\cite{YooandRadmorebrief}. The same
phenomenon was shown to appear in systems with a continuum of
levels~\cite{Deng} and in quantum systems interacting with
quantized fields~\cite{Sanchez1}. In all of these cases the
mechanism is almost the same: population trapping occurs when
there exists a superposition of states which decouples from the
other states, so that its population is constant in time. In this
case, the various transition channels corresponding to the states
in the superposition interfere destructively, canceling the
decay~\cite{Deng}. An analogous phenomenon has been singled out in
the dissipative dynamics of an atom interacting with a structured
reservoir. Indeed, for a two-level atom interacting with the
quantized e.m. modes of a photonic-bandgap (PBG)
material~\cite{john87,angelakisreview}, the atomic population can
be partially trapped in the excited state, when the atomic Bohr
frequency is near the edge of the
gap~\cite{john90,john94,quang97,lambropoulosreview}. In this case
the trapping is due to the formation of two atom-photon dressed
states, one of which, due to strong vacuum Rabi splitting, is
protected against decay because the energy of the relevant
transition to the ground state lies inside the gap, as explained
in Ref.~\cite{lambropoulosreview}. This has led to an extensive
theoretical and experimental analysis of the physics of cavities
inside PBG materials interacting either with real atoms~\cite{Lev}
or quantum dots~\cite{lflorescu,reithmaierandyoshie}.

As discussed in Ref.~\cite{lambropoulosreview}, what was missing
until some years ago was the inclusion of cavity losses in the
study of the dynamics of these systems. The cavity losses can be
thought as due to an imperfect realization of the experimental
setup for the study of these systems. More precisely they arise
from the imperfect alignment of the auxiliary waveguides which
couple to the cavities according to the designed
configurations~\cite{lflorescu}. Recently the effect of cavity
losses has been studied using a phenomenological
dissipator~\cite{lflorescu,Lev}, identical to the one used to
describe the losses in cavity quantum electrodynamics
(CQED)~\cite{haroche}. However, one may wonder if and in which
limits this phenomenological model is consistent with the
description of a set of waveguides in the PBG material as a
bosonic reservoir, as done in Ref.~\cite{longhi2006}.

The scope of this paper is to provide a microscopic derivation of
the master equation when we deal with CQED involving PBG materials
with cavity losses, in the framework of an appropriate
non-Markovian theory. We will show that the dynamics of the system
shows population trapping in the atomic excited state. This effect
arises because the decay rates appearing in the microscopic
dissipator are different.

The system we study consists of a two-level atom interacting with
a cavity mode, where the cavity is coupled to a bosonic
environment. The interaction between the atom and the system is
described, at resonance and in units of $\hbar$, by the JC
Hamiltonian $H_{JC}=(\omega_0/2)\sigma_z+\omega_0\,a^\dag
 a+\Omega\left(a\sigma_++a^\dag\sigma_-\right)$, where  $a^\dag$ ($a$) is the creation (annihilation) operator of
the mode, $\sigma_-=\left|g\right>\left<e\right|$,
$\sigma_+=\left|e\right>\left<g\right|$, and $\sigma_z =
\left|e\rangle \langle e \right| - \left|g \rangle \langle g
\right|$, while $\left|g\right>$ and $\left|e\right>$ denote the
atomic ground and excited state respectively~\cite{JC}. This model
is valid as long as $\Omega\ll\omega_0$, so that one can neglect
the counter-rotating terms $a\sigma_-$ and $a^\dag\sigma_+$. The
cavity mode interacts with a bosonic reservoir, with Hamiltonian
$H_R=\sum_k\omega_kb_k^\dag b_k$, through the interaction
Hamiltonian $H_{\rm int}=(a+a^\dag)\sum_kg_k(b_k+b_k^\dag)$, which
has the advantage of being treatable straightforwardly in the
time-convolutionless formalism we are exploiting in this paper.
Since the reservoir causing cavity losses is immersed in the PBG
material, we expect its spectrum to be non-flat: to be rigorous,
the master equation must be derived in the framework of a
non-Markovian theory. Using the second order of the
time-convolutionless (TCL)
expansion~\cite{petruccionebook,noteonTCL}, and neglecting the
atomic spontaneous emission and the Lamb shifts, the master
equation, in the strong atom-cavity coupling regime, is equal to
the Markovian one under the rotating wave
approximation~\cite{Scala,Scala2}, with the important difference
that now the decay rates are time-dependent, indicating
non-Markovian behavior. In this paper we focus on the case of one
initial excitation and we consider a reservoir at zero
temperature. In this case, the non-Markovian master equation for
the atom-cavity system density operator $\rho$ is:
\begin{eqnarray}
 \dot{\rho}(t)&=&-i\left[H_{JC},\rho\right]\nonumber\\
\nonumber\\ &+&\gamma\left(\omega_0+\Omega,t\right)
  \left(\frac{1}{2}\left|E_0\right>\left<E_{1,+}\right|\rho(t)\left|E_{1,+}\right>\left<E_0\right|\right.\nonumber\\
  &-&\left.\frac{1}{4}\left\{\left|E_{1,+}\right>\left<E_{1,+}\right|,\rho(t)\right\}\right)\nonumber\\
  &+&\gamma\left(\omega_0-\Omega,t\right)
  \left(\frac{1}{2}\left|E_0\right>\left<E_{1,-}\right|\rho(t)\left|E_{1,-}\right>\left<E_0\right|\right.\nonumber\\
\nonumber\\
&-&\left.\frac{1}{4}\left\{\left|E_{1,-}\right>\left<E_{1,-}\right|,\rho(t)\right\}\right),
\label{masterequation}
\end{eqnarray}
where
$\left|E_{1,\pm}\right>=(1/\sqrt{2})(\left|1,g\right>\pm\left|0,e\right>)$
are the eigenstates of $H_{JC}$ with one total excitation, with
energy $\omega_0/2\pm\Omega$, and
$\left|E_0\right>=\left|0,g\right>$ is the ground state, with
energy $-\omega_0/2$. The time-dependent decay rates for
$\left|E_{1,-}\right>$ and $\left|E_{1,+}\right>$ are
$\gamma(\omega_0-\Omega,t)$ and $\gamma(\omega_0+\Omega,t)$
respectively.

If the system starts from the state $\left|0,e\right>$, i.e., if
the atom is initially excited and the cavity is initially empty,
from Eq.~(\ref{masterequation}) one can obtain the system density
operator at all times:
\begin{eqnarray}\label{rhorabinonmarkovian}
 &&\rho(t)=\left(1-\frac{1}{2}\mathrm{e}^{-\frac{I_-(t)}{2}}-\frac{1}{2}\mathrm{e}^{-\frac{I_+(t)}{2}}\right)
 \left|E_0\right>\left<E_0\right|\nonumber\\
\nonumber\\
 &&+\frac{1}{2}\mathrm{e}^{-\frac{I_-(t)}{2}}\left|E_{1,-}\right>\left<E_{1,-}\right|+
 \frac{1}{2}\mathrm{e}^{-\frac{I_+(t)}{2}}\left|E_{1,+}\right>\left<E_{1,+}\right|\nonumber\\
\nonumber\\
 &&-\frac{1}{2}\mathrm{e}^{-\frac{I_-(t)+I_+(t)}{4}}\left(\mathrm{e}^{2i\Omega t}\left|E_{1,-}\right>\left<E_{1,+}\right|+
\mbox{h.c.}\right),
\end{eqnarray}
where $I_\pm(t)=\int_0^t\gamma(\omega_0\pm\Omega,t')dt'$. From
Eq.~(\ref{rhorabinonmarkovian}) it is possible to compute all the
populations that we will show in the following. Below we will
study the behavior of the non-Markovian time-dependent rates
$\gamma(\omega_0\pm\Omega,t)$, which, through the quantities
$I_\pm(t)$, lead to population trapping.

As a model of environment at zero temperature with non-flat
spectrum, we consider the Lorentzian
distribution~\cite{petruccionebook}:
\begin{equation}\label{lorentziandensity}
  J(\omega)=\frac{1}{2\pi}\frac{\alpha\lambda^2}{(\omega_1-\omega)^2+\lambda^2},
\end{equation}
where $\alpha$ is the system-environment coupling strength, and
$\lambda$ is the width of the distribution, describing also the
inverse of the reservoir memory time. The case of Lorentzian
spectrum is analytically treatable, while capturing important
features of the non-Markovian dynamics we are interested in, i.e.
the time-dependence of the decay rates and their different
stationary values. We consider the case in which the spectrum is
peaked on the frequency of the state $\left|E_{1,-}\right>$, i.e.,
$\omega_1=\omega_0-\Omega$, where $\omega_0$ is the atomic Bohr
frequency and $\Omega$ is the Rabi splitting due to the JC
interaction. The rate $\gamma(\omega,t)$ for a generic transition
with Bohr frequency $\omega$ is equal to
$\gamma(\omega,t)=2\mathrm{Re}\left\{\Gamma(\omega,t)\right\}$,
where $ \Gamma(\omega,t)$ is related to the spectral density
$J(\omega)$ through the relation:
\begin{equation}\label{Gamma_t_dep}
 \Gamma(\omega,t)=\int_0^td\tau\int_{-\infty}^{+\infty}d\omega'\mathrm{e}^{i(\omega-\omega')\tau}J(\omega').
\end{equation}

By performing first the integral with respect to $\tau$ and then
calculating the remaining integral by means of the method of the
residues, we obtain the following expression for the decay rate
$\gamma(\omega,t)$:
\begin{eqnarray}\label{gamma_tdep}
 &&\gamma(\omega,t)=\frac{\alpha\lambda^2}{(\omega_1-\omega)^2+\lambda^2}\left\{1+\left[\frac{\omega_1-\omega}{\lambda}\sin(\omega_1-\omega)t
 \right.\right.\nonumber\\
 \nonumber\\
 &&
 -\left.\cos(\omega_1-\omega)t\Big]\mathrm{e}^{-\lambda t}\right\}.
\end{eqnarray}

In particular, for $\omega_1=\omega_0-\Omega$ and substituting
$\omega=\omega_0\pm\Omega$, we obtain the decay rates for the two
dressed states $\left|E_{1,\pm}\right>$:
\begin{equation}\label{gamma_E-}
 \gamma(\omega_0-\Omega,t)=\alpha\left(1-\mathrm{e}^{-\lambda t}\right),
\end{equation}
for $\left|E_{1,-}\right>$ and
\begin{eqnarray}\label{gamma_E+}
 &&\gamma(\omega_0+\Omega,t)=\frac{\alpha\lambda^2}{(2\Omega)^2+\lambda^2}
 \left\{1+\left[\frac{2\Omega}{\lambda}\sin2\Omega
 t\right.\right.\nonumber\\
 \nonumber\\
 &&-\left.\cos2\Omega t\Big]\mathrm{e}^{-\lambda t}\right\},
\end{eqnarray}
for $\left|E_{1,+}\right>$.

\begin{figure}
 \begin{center}
     \includegraphics[ width=0.46\textwidth, height=0.28\textwidth ]{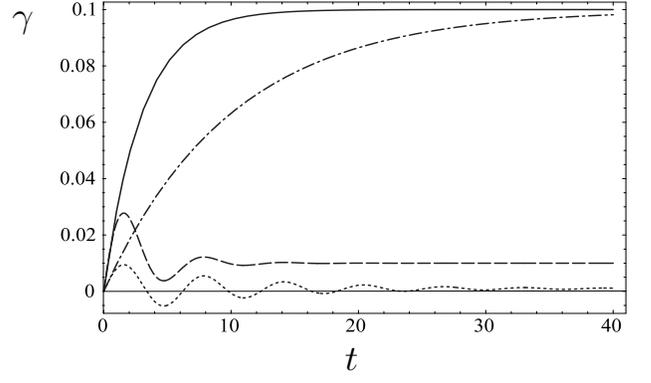}
    \caption{Time-dependent decay rates (in units of $2\Omega$) as a function of $t$ (in units of $(2\Omega)^{-1}$) for the examples we
    are considering: ({\em i}) $\lambda=2\Omega/3$, rates for $\left|E_{1,-}\right>$ (solid line) and for $\left|E_{1,+}\right>$ (dashed line); ({\em ii})
    $\lambda=2\Omega/\sqrt{99}$, rates for $\left|E_{1,-}\right>$ (dashed-dotted line) and
    for $\left|E_{1,+}\right>$ (dotted line).}\label{nonmarkratesbis}
 \end{center}
\end{figure}

From Eqs.~(\ref{gamma_tdep})-(\ref{gamma_E+}) we clearly see the
general behavior of the time-dependent rates: all the rates
$\gamma(\omega,t)$ are zero at $t=0$, then they increase in time,
till they reach stationary values for $t\gg\lambda^{-1}$. These
stationary values are proportional to $J(\omega)$,~i.e.,~they are
equal to the rates one obtains from a Markovian theory. For these
reasons, the quantity $\lambda^{-1}$ can be seen as the memory
time of the system-reservoir interaction and non-Markovian effects
are expected to occur for times shorter than  $\lambda^{-1}$ .

Figure~\ref{nonmarkratesbis} shows the decay rates for two cases,
corresponding respectively to $\lambda/(2\Omega)=1/3$ and
$\lambda/(2\Omega)=1/\sqrt{99}$, with $\alpha/(2\Omega)=1/10$ for
both cases. The condition on $\alpha$ assures the strong coupling
regime, since the asymptotic decay rate of the state
$\left|E_{1,-}\right>$ is $\gamma(\omega_0-\Omega,
\infty)=\alpha\ll2\Omega$   in both cases. As for the asymptotic
decay rate of the state $\left|E_{1,+}\right>$, in the first case
$\lambda/(2\Omega)=1/3$ corresponds to $\gamma(\omega_0+\Omega,
\infty)=\gamma(\omega_0-\Omega,
\infty)/10$, while in the second case 
$\lambda/(2\Omega)=1/\sqrt{99}$ corresponds to
$\gamma(\omega_0+\Omega, \infty)=\gamma(\omega_0-\Omega,
\infty)/100$. Note that in the case
$\lambda/(2\Omega)=1/\sqrt{99}$ the decay rate for
$\left|E_{1,+}\right>$ also reaches negative values for short
times: this is a typical feature of non-Markovian decay rates when
the correlation time of the reservoir becomes
large~\cite{petruccionebook}, and recently this has been connected
to memory effects restoring coherence, for short times, within a
quantum jump scheme~\cite{jumpsnegative}.

By looking at the population $P_{0,g}$ of the system ground state
$\left|0,g\right>$, one can monitor the loss of energy from the
atom-cavity system. 
\begin{figure}
 \begin{center}
     \includegraphics[ width=0.46\textwidth, height=0.28\textwidth ]{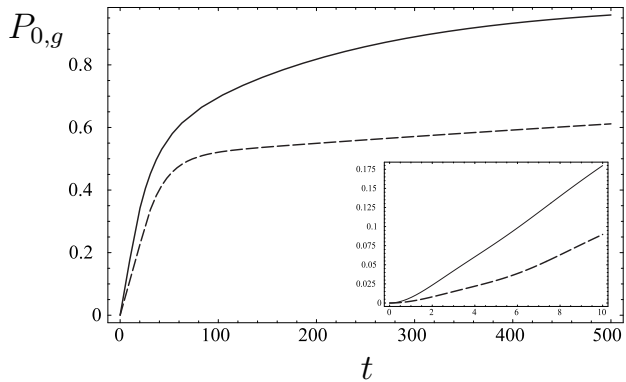}
    \caption{Population of the ground state of the atom-cavity system,
    as a function of $t$ (in units of $(2\Omega)^{-1}$) for the cases $\lambda=2\Omega/3$ (solid line) and $\lambda=2\Omega/\sqrt{99}$
    (dashed line). The inset shows the short-time dynamics.}\label{nonmarkogog}
 \end{center}
\end{figure}
Figure~\ref{nonmarkogog} shows the time evolution of $P_{0,g}$ for
the two cases we are analyzing. Let us first consider the case
$\lambda/(2\Omega)=1/3$ (solid line). The long-time dynamics is
well described by a decay law which is a sum of two 
exponentials, one with rate $\gamma(\omega_0-\Omega, \infty)$ and
another one with rate $\gamma(\omega_0+\Omega, \infty)$, slower
than the first one. When probing the population of the ground
state, a proper analysis of the signal should allow to point out
this feature and to distinguish it from the purely exponential
decay predicted by the phenomenological dissipator for cavity
losses in strong coupling. The inset in Fig.~\ref{nonmarkogog}
shows the short-time evolution of the population under scrutiny,
where the non-Markovian effects are evident. For very short times
the rates start from zero and then increase linearly in time, as
can be seen by expanding in power series of $t$ the rates in
Eqs.~(\ref{gamma_tdep})-(\ref{gamma_E+}). Hence the  short-time
behavior of the population is quadratic in time, in agreement with
the predictions of the non-Markovian
theory~\cite{petruccionebook}. As time increases, signatures of
the oscillations of the rate $\gamma(\omega_0+\Omega, t)$ may
appear in the dynamics. For $\lambda/(2\Omega)=1/3$ none of them
is visible. For $\lambda/2\Omega=1/\sqrt{99}$ (dashed line in
Fig.~\ref{nonmarkogog}) the main features of the dynamics remain
unchanged but the short-time dynamics shows, along with the
initial quadratic behavior, also slight signatures of the
oscillations in the rate, consisting in a non-quadratic increase
in the ground state population, with non-monotonic derivative.
Such signatures are now visible both because the memory time of
the reservoir is much longer than in the first case and because
the decay rate of the state $\left|E_{1,+}\right>$ also reaches
negative values for some time intervals (see the dotted line in
Fig.~\ref{nonmarkratesbis}). 

In the last case, the large difference between the two asymptotic
decay rates leads to an interesting picture of the long-time 
dynamics. Indeed the ground state population first reaches the
value of about $50\%$ and stays close to this value for a very
long time, then the system starts to decay again and the ground
state population eventually goes to $1$ (this restarting of the
decay is not shown in Fig.~\ref{nonmarkogog}). It is as if the
second decay channel is turned on after a time of the order of
$\gamma(\omega_0+\Omega,\infty)^{-1}$. Before this time the decay
of $\left|E_{1,+}\right>$ is almost completely inhibited, which is
equivalent to the ideal situation wherein one of the two decay
channels due to cavity losses is completely closed. This is at the
origin of the population trapping due to cavity losses, as we are
going to see.

\begin{figure}
\begin{center}
     \includegraphics[ width=0.46\textwidth, height=0.28\textwidth ]{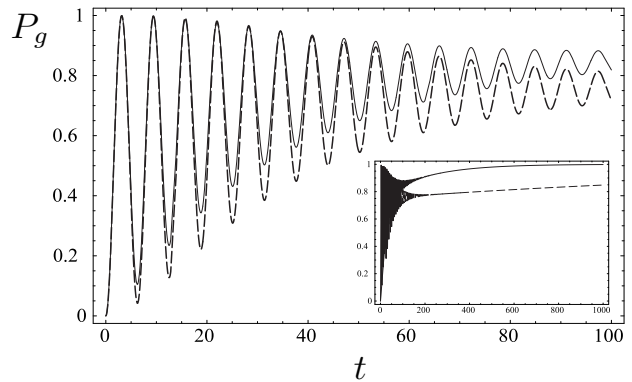}
    \caption{Population of the atomic ground state, as a function of $t$ (in units of $(2\Omega)^{-1}$), for the cases $\lambda=2\Omega/3$ (solid line)
    and $\lambda=2\Omega/\sqrt{99}$ (dashed line). The inset shows the very long time
    behavior.
    }\label{nonmarkpopgnocolor}
 \end{center}
\end{figure}

Figure~\ref{nonmarkpopgnocolor} shows the population of the atomic
ground state $\left|g\right>$. This is the quantity which is
usually measured in standard CQED experiments after the atom has
left the cavity~\cite{haroche}. As in standard CQED the population
exhibits Rabi oscillations, but the dynamics of the decay of the
oscillations is quite different. Since the two decay channels have
different asymptotic rates, the state $\left|E_{1,-}\right>$
decays before the state $\left|E_{1,+}\right>$. Therefore there is
a time interval, after the decay of $\left|E_{1,-}\right>$ and
before the complete decay of $\left|E_{1,+}\right>$, wherein the
Rabi oscillations are washed away by cavity losses, but some
excitation is still present in the atom-cavity system. This is
reflected in the fact that some population is trapped in the
excited state of the atom for very long time, as shown in
Fig.~\ref{nonmarkpopgnocolor}. Comparing the cases
$\lambda/(2\Omega)=1/3$ and $\lambda/2\Omega=1/\sqrt{99}$, we can
see that in both cases the phenomenon of trapping is clearly
visible. The difference is in the amount of trapping and in the
length of the time interval wherein the trapping is present. The
trapping time is longer in the second case, since the decay of
$\left|E_{1,+}\right>$ is much slower than in the first one.
Anyway in both cases the trapping occurs for a time interval much
longer than a Rabi period. The amount of trapping in the atomic
excited state, in the first case, is a bit less than $20\%$ while
in the second case it is close to $25\%$. It is clear that the
amount of population trapping increases when the ratio between the
two rates becomes smaller. The values we have shown in the latter
case are very close to the limit values for population trapping
due to cavity losses. Indeed the limiting case is the one wherein
one rate is zero. In this case one of the two dressed states does
not decay and the long-time population of the system ground state
$\left|0,g\right>$ is $50\%$. The remaining population corresponds
to the non-decaying dressed state, which is a superposition of the
states $\left|0,e\right>$ and $\left|1,g\right>$ with equal
weights. This means that the limit value achievable for the
population trapped in the atomic excited state is equal to $25\%$.

In conclusion, we have shown that a microscopic derivation of the
non-Markovian master equation for the JC model when the cavity
interacts with an external environment, allows one to predict
population trapping due to cavity losses when the spectrum of the
environment is not flat. It is important to note that the
population trapping is a feature of the dynamics arising when the
rates of the two dressed states are different. This means that
even a non-Markovian extension of the phenomenological model used
to describe cavity losses for the JC model, which would contain
only one single time-dependent decay rate, could not predict this
effect. Only a microscopic derivation of the master equation for
the JC model does allow to correctly describe cavity losses for a
non-flat spectrum of the environment and to predict the occurrence
of population trapping in the dressed state,
$\left|E_{1,+}\right>$. This state is protected against cavity
losses for a very long time, and its photon-atom entanglement
could be exploited,~e.g.,~to perform experiments on the violation
of Bell's inequality. The long lifetime could indeed allow one to
overcome the locality loophole, making it possible to send the
atom away from the cavity far enough to be outside of the cavity
lightcone~\cite{localityloophole}. A very important point is that
our approach paves the way to the inclusion of cavity losses in
the study of more complex forms of CQED in PBG materials. In
particular our microscopic dissipator could be included in some
schemes of quantum feedback control of the quantum dynamics of the
atom-cavity system~\cite{mabuchifeedback} and in the study of the
physics of arrays of cavities. These latter systems  have been
recently exploited to simulate, through photon-atom dressed
states, important models of condensed matter
physics~\cite{array_cavity}.

M.S. acknowledges financial support from CIMO for his stay in
Turku during the period January-April 2007 and wishes to thank all
the Quantum Optics Group members of the University of Turku for
their kind hospitality. S.M., J.P. and K.-A.S. acknowledge
financial support from the Academy of Finland (projects 108699,
115982, 115682), the Magnus Ehrnrooth Foundation and the
V\"{a}is\"{a}l\"{a} Foundation.


\begin{thebibliography}{99}
\bibitem{YooandRadmorebrief} H.I. Yoo and J.H. Eberly, Phys. Rep. {\bf 118}, 239
(1985) and references therein; P.M. Radmore, Phys. Rev. A {\bf
26}, 2252 (1982).
\bibitem{Deng} Z. Deng and J.H. Eberly, Phys. Rev. A {\bf 34},
2492 (1986).
\bibitem{Sanchez1} J.I. Cirac and L.L. S\'anchez-Soto, Phys. Rev.
A {\bf 42}, 2851 (1990); {\em ibid.} {\bf 44}, 3317 (1991); G.S.
Agarwal, Phys. Rev. Lett. {\bf 71}, 1351 (1993).
\bibitem{john87} S. John, Phys. Rev. Lett. {\bf 58}, 2486 (1987).
\bibitem{angelakisreview} D.G. Angelakis {\em et al.}, Contemp.
Phys. {\bf 45}, 303 (2004). 
\bibitem{john90} S. John and J. Wang, Phys. Rev. Lett. {\bf 64},
2418 (1990).
\bibitem{john94} S. John and T. Quang, Phys. Rev. A {\bf 50},
1764 (1994).
\bibitem{quang97} T. Quang {\em et al.}, Phys. Rev. Lett. {\bf
79}, 5238 (1997).
\bibitem{lambropoulosreview} P. Lambropoulos {\em et al.}, Rep.
Prog. Phys. {\bf 63}, 455 (2000).
\bibitem{Lev} B. Lev {\em et al.}, Nanotechnology {\bf 15}, S556
(2004).
\bibitem{lflorescu} L. Florescu {\em et al.}, Phys. Rev. A {\bf
69}, 013816 (2004).
\bibitem{reithmaierandyoshie} J.P. Reithmaier {\em et al.}, Nature {\bf
432}, 197 (2004); T. Yoshie {\em et al.}, Nature {\bf 432}, 200
(2004); D. Englund {\em et al.}, Nature {\bf 450}, 857 (2007); K.
Srinivasan and O. Painter, Nature {\bf 450}, 862 (2007).
\bibitem{haroche} S. Haroche {\it et al.}, in {\em Fundamental Systems in Quantum
Optics}, ed. by J. Dalibard, J.-M. Raimond and J. Zinn-Justin
(North Holland, Amsterdam, 1992).
\bibitem{longhi2006} S. Longhi, Phys. Rev. A {\bf 74}, 063826
(2006).
\bibitem{JC} E.T. Jaynes and F.W. Cummings, Proc. IEEE {\bf
51}, 89 (1963).
\bibitem{petruccionebook}  H.-P. Breuer and F. Petruccione, {\it The Theory of Open
Quantum Systems}  (Oxford University Press, Oxford, 2002).

\bibitem{noteonTCL} The TCL formalism consists in a perturbative
expansion, in the system-enivironment coupling constant, of an
exact master equation for the open system under scrutiny. This
approach is valid as long as such a coupling is weak enough, which
in our case corresponds to the condition
$\gamma(\omega_0\pm\Omega)\ll2\Omega$ \cite{Scala}.

\bibitem{Scala} M. Scala {\em et al.}, Phys. Rev. A {\bf 75}, 013811
(2007).

\bibitem{Scala2} M. Scala {\em et al.}, J. Phys. A: Math. Theor.
{\bf 40}, 14527 (2007).

\bibitem{jumpsnegative} J. Piilo {\em et al.}, arXiv:0706.4438v2.
\bibitem{localityloophole} R. Pearle, Phys. Rev. D {\bf 2}, 1418
(1970).
\bibitem{mabuchifeedback} J.E. Reiner {\em et al.}, Phys. Rev. A
{\bf 67}, 042106 (2003); M.A. Armen and H. Mabuchi, Phys. Rev. A
{\bf 73}, 063801 (2006); D.A. Steck {\em et al.}, Phys. Rev. A
{\bf 74}, 012322 (2006).
\bibitem{array_cavity} M.J. Hartmann and M.B. Plenio, Nature Physics {\bf 2}, 849 (2006);
A.D. Greentree {\em et al.}, Nature Physics {\bf 2}, 856 (2006);
D.G. Angelakis {\em et al.}, Phys. Rev. A {\bf 76}, 031805(R)
(2007); M.J. Hartmann and M.B. Plenio, Phys. Rev. Lett. {\bf 99},
103601 (2007); D. Rossini and R. Fazio, Phys. Rev. Lett. {\bf 99},
186401 (2007).

\end{thebibliography}
\end{document}